\documentclass[reprint, aps, onecolumn, amsmath, amssymb]{revtex4-2}
\usepackage{graphicx}
\usepackage{newtxtext}
\usepackage{newtxmath}
\usepackage{natbib}
\usepackage{hyperref}
\hypersetup{
    colorlinks = true,
    urlcolor   = blue,
    citecolor  = black,
}
\usepackage[separate-uncertainty = true,multi-part-units=single]{siunitx}
\usepackage{mathtools}
\usepackage{gensymb}
\usepackage{amsmath}
\usepackage{physics}
\usepackage{stmaryrd}
\usepackage{dcolumn}
\usepackage{bm}

\newcommand{\RomanNumeralCaps}[1]
\linenumbers

\usepackage[nameinlink,noabbrev]{cleveref}
\crefformat{section}{\S#2#1#3}%
\crefformat{subsection}{\S#2#1#3}
\crefformat{subsubsection}{\S#2#1#3}
\crefformat{figure}{#2Fig.~#1#3}
\crefmultiformat{figure}{Figs.~#2#1#3}{ and~#2#1#3}{, #2#1#3}{ and~#2#1#3}

\begin{document}
	
	\preprint{APS/123-QED}
	
	\title{On the oscillatory dynamics of a Saffman--Taylor finger with a bubble at its tip\\}
	
	\author{Jack Lawless}
	\email{jack.lawless@manchester.ac.uk}
	\author{Andrew L. Hazel}%
	\author{Anne Juel}
	\affiliation{Manchester Centre for Nonlinear Dynamics, The University of Manchester, Oxford Road, Manchester, M13 9PL, UK\\}
	
	\date{\today}


\begin{abstract}
 The complex behaviour of air-liquid interfaces driven into Hele-Shaw channels at high speeds could arise from oscillatory dynamics; yet, both the physical and dynamical mechanisms that lead to interfacial oscillations remain unclear.
 We extend the experiments by Couder \textit{et. al.} (\textit{Phys. Rev. A}, vol. 34, 1986, p. 5175) to present a systematic investigation of the dynamics
that result when a small air bubble is placed at the tip of a steadily propagating air finger in a Hele-Shaw channel. The system can exhibit steady and oscillatory behaviour, and we show that these different behaviours each occur in well-defined regions of the phase space defined by flow rate and bubble size. For sufficiently large flow rates, periodic finger oscillations give way to disordered dynamics characterised by an irregular meandering of the finger's tip. 

We demonstrate that at a fixed flow rate, the oscillations commence when the bubble size is increased sufficiently so that the decreased in-plane curvature of the bubble tip matches the in-plane curvature of the finger tip. The equality between the two in-plane curvatures causes the axial pressure gradient across the bubble, which drives the finger, to vanish, thus rendering the finger susceptible to lateral perturbations. Differing timescales for finger and bubble restoral under perturbation allow sustained oscillations to develop in the finger-bubble system. The oscillations cease when the bubble is sufficiently large that it can act as the tip of a compound finger. The disordered dynamics at high flow rates are consistent with the transient exploration of unstable periodic states, which suggests that similar dynamics may underlie the observed disordered dynamics in viscous fingering.
\end{abstract}
\maketitle
\section{Introduction}
\label{sec:introduction}

\begin{figure}
\includegraphics[width=\textwidth, clip]{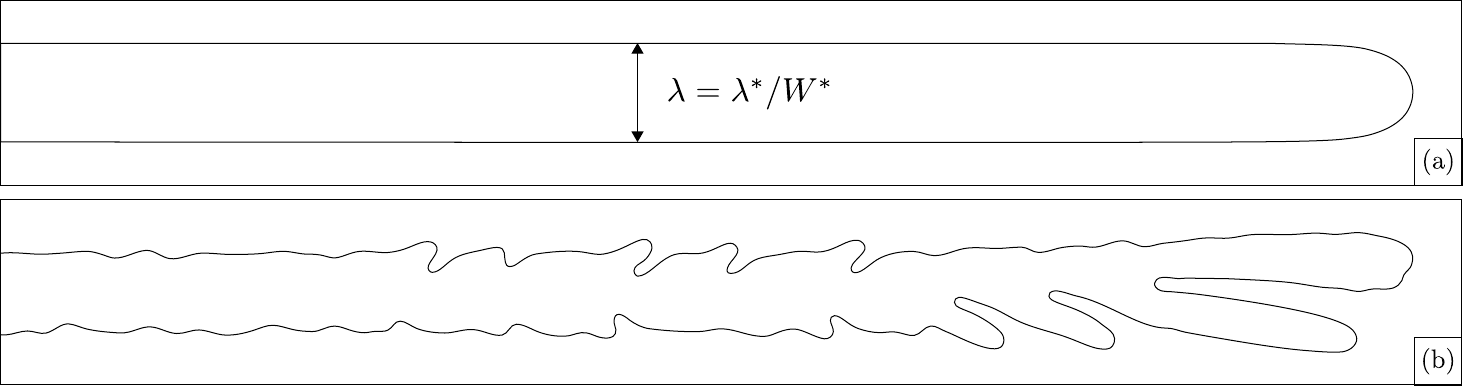}
\caption{The typical fingering patterns that develop when a constant
  flux of air invades a viscous liquid-filled Hele-Shaw at two
  different values of the dimensionless parameter $1/B = 12 \:
  \alpha^2 \: \mathrm{Ca}$, where $\alpha = W^* / H^*$ is the
  channel's cross-sectional aspect ratio and $\mathrm{Ca} = \mu U^* /
  \sigma$ is a capillary number. Here, $\mu$ is the dynamic viscosity
  of the liquid, $\sigma$ is the surface tension at the air-liquid interface and $U$* is the finger's propagation speed. (a) The single, steadily propagating and symmetric Saffman--Taylor finger that develops for $1/B = 1200$. The channel's width is $W^* =$ \SI{40.0 (1)}{\milli\metre} and its depth is $H^* =$ \SI{1.0 (1)}{\milli\metre}. The liquid's dynamic viscosity is $\mu =$ \SI{0.019}{\pascal \second}. The finger width $\lambda^{*}$ is also indicated in the figure. (b) The disordered finger that develops for $1/B = 15,000$. The channel's width is $W^* =$ \SI{60.0 (1)}{\milli\metre} and its depth is $H^* =$ \SI{1.0 (1)}{\milli\metre}. The liquid's dynamic viscosity is $\mu =$ \SI{0.095}{\pascal \second}. The surface tension at the air-liquid interface is $\sigma =$ \SI{21}{\milli \newton \per \metre} in both cases. This figure has been reproduced from \citet{osc_formations} with permission.}
\label{fig:ST_vs_disordered}
\end{figure}

The displacement of a more-viscous fluid by a less-viscous fluid inside a quasi-two-dimensional (Hele-Shaw) channel gives rise to a fascinating pattern-forming phenomenon known as viscous fingering \citep{perspectives}, see Fig.~\ref{fig:ST_vs_disordered}. The patterns are formed by the growth of intricate finger-like protrusions at the morphologically unstable interface between the two fluids, and the detailed dynamics depends on the channel's geometry, fluid miscibility and propagation speed of the interface \citep{injection, viscosity_ratio, zheng, Juel2018}. The allure of these patterns stems from their irregularity, which is a direct consequence of the system's propensity to exhibit ``disordered'' growth dynamics \citep{Paterson_1981, Chen_1989, arneodo, Kopf_Sill}. Whilst attempts have been made to characterise these patterns from a phenomenological perspective, with morphological descriptions such as their fractal dimensions \citep{Zhang, Swinney, beeson_jones} being used to quantify their complexity, the underlying drivers of the system's disordered dynamics have remained poorly understood from both a dynamical systems and a physical perspective \citep{Jack2025}. We are particularly interested in periodic dynamics because unstable periodic states can be the ``building blocks'' of disordered dynamics, as highlighted in studies of the subcritical transition to turbulence in linearly stable wall-bounded shear flows, where analysis of unstable periodic states has led to significant progress in understanding the time-averaged statistical properties of turbulent flows \citep{eckhardt, duguet, budanur, Kawahara, Cvitanovic, Suri}. We hypothesise that the complex time-dependent behaviour of viscous fingers is orchestrated by (dynamically disconnected) unstable periodic states, which motivates the present experimental investigation into periodic propagation dynamics.

Previous experiments have shown that Saffman--Taylor fingers become increasingly sensitive to perturbations as the driving parameter is increased \citep{tabeling, Chevalier2}, and the nonlinear stability analysis performed by \cite{bensimon} indicates that the amplitude of perturbation required for instability decreases exponentially with increasing values of this driving parameter. Perturbations of the finger tip can lead to selection of ``anomalous'' steadily propagating fingers: introduction of a thin thread \citep{Zocchi} leads to narrower symmetric or asymmetric fingers; the presence of a bubble at the finger tip \citep{couder} can also lead to narrower fingers, a finding later confirmed theoretically \citep{honglanger, combescotdombre}; etching grooves or a microscopic lattice into a channel boundary \citep{rabaud, lattice} also leads to similar long-term behaviours. 

 In this paper, we revisit the experiments originally carried out by \citet{couder},
 in which the Saffman--Taylor finger was perturbed by placing a bubble
 at its tip. The bubble provides a unique means of perturbing the tip of the finger because, unlike channel geometry perturbations, this perturbation remains localised to the tip of the finger as it propagates. The original experiments were
 motivated by the so-called finger selection problem, which concerns the associated depth-averaged theory's lack of a
 selection mechanism in the absence of surface tension for the half-width finger that is observed in experiments \cite{Saffman1958}. However, they also uncovered the propagation of dynamically intriguing fingers with either oscillating or pulsating tips.

The preliminary observations of periodic oscillations by \citet{couder} have since been complemented by similar findings in other geometrically perturbed systems. For example, the
depth-perturbed system that was studied by \citet{hazel_pailha}
contains a multitude of stable and unstable periodic states \citep{pailha, Keeler2019, Lawless}. The experiments by \citet{cuttle} inside elasto-rigid Hele-Shaw channels, where the channel's upper boundary was replaced by a deformable elastic membrane, also revealed
complex unsteady dynamics; and a host of periodic states were found by \citet{Fontana_2023} in a model of that system. The most
comprehensive study of periodic dynamics was carried out by \citet{osc_formations}, who initiated stable oscillations by remotely
perturbing the finger's tip with neighbouring bubbles. The studies clearly demonstrate that the finger has a propensity to exhibit oscillations when controlled perturbations are applied, but these perturbations typically modify the system's bifurcation structure and it is indeed possible that the oscillatory states do not exist in the unperturbed system, although there is theoretical evidence for a countably infinite number of unstable steadily propagating states \citep{romero, romero2, gardiner}. 


 Our principal aim in the present paper is to examine the mechanisms that lead to the observed periodic oscillations by systematically varying the bubble's size and flow rate in experiments 
 in order to establish empirical criteria for when the steadily propagating finger destabilises. For sufficiently large flow rates, our experiments also reveal novel disordered finger dynamics, which are consistent with the transient exploration of multiple unstable periodic states.

We have organised the paper as follows. Firstly, the experimental set-up and methods that are used to generate bubbles are described in Sec.~\ref{sec:setup}. Because the underlying physical mechanism that leads to periodic dynamics requires prior knowledge of the Saffman--Taylor finger, we describe this in Sec.~\ref{sec:ST_finger}. We establish the region of the system's parameter space in which periodic oscillations and disordered dynamics occur by systematically varying the bubble's size and flow rate, and constructing a phase diagram in Sec.~\ref{phase_diagram}. The underlying physical mechanism that leads to periodic oscillations is described in Sec.~\ref{oscillations}. Finally, we discuss the significance of our findings and avenues for future work in Sec.~\ref{conclusion}.

\section{Experimental methods}

\subsection{Experimental set-up}
\label{sec:setup}

\begin{figure}
\includegraphics[width=\textwidth, clip]{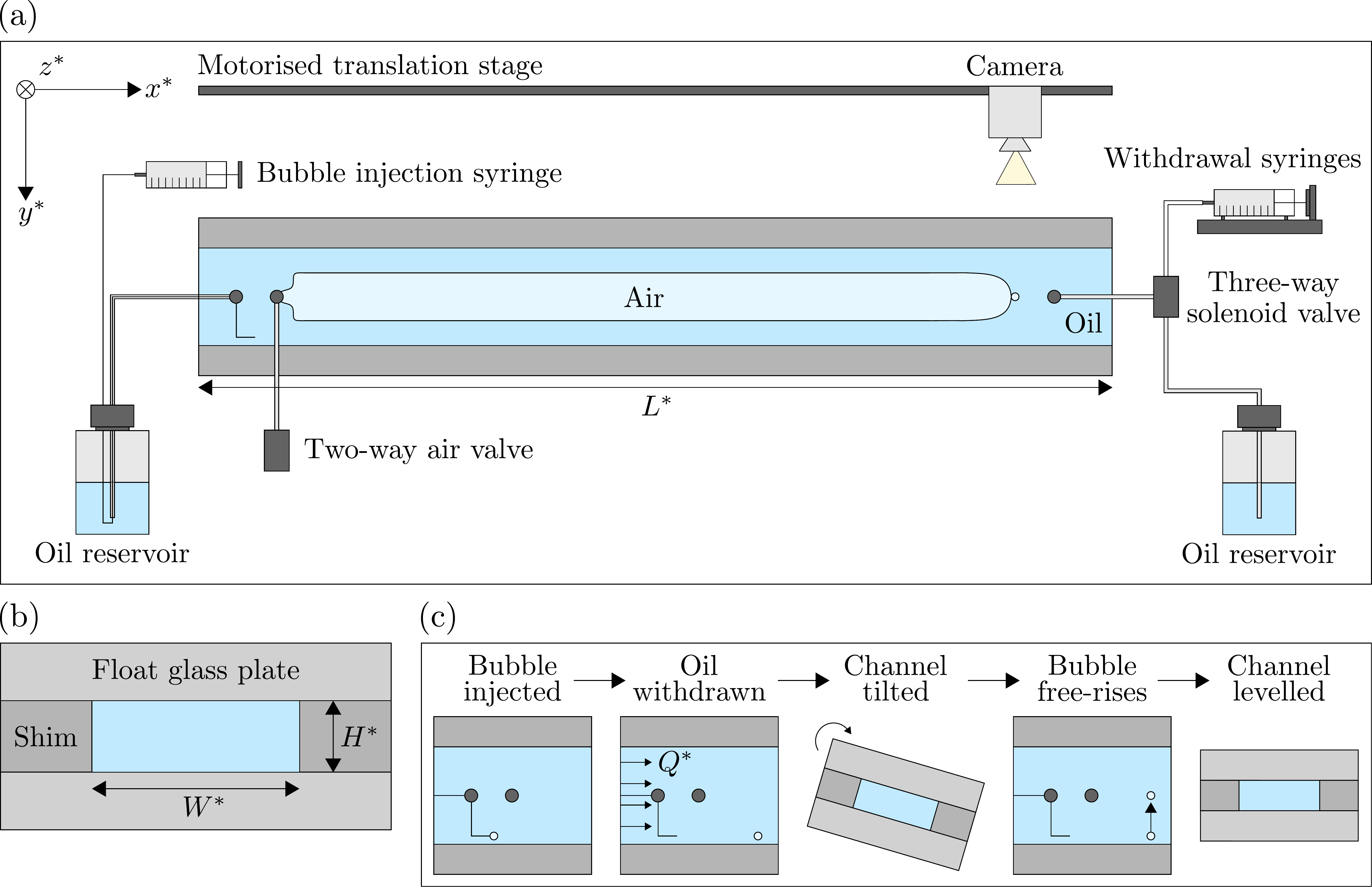}
\caption{(a) The horizontally levelled rectangular Hele--Shaw
  channel. The channel's width is $W^* =$ \SI{40.0 (1)}{\milli\metre}
  and its depth is $H^* =$ \SI{1.0 (1)}{\milli\metre}. The channel
  outlet is connected to a network of syringe pumps and an external
  oil reservoir by a three-way solenoid valve. The channel inlet is
  connected to an external oil reservoir. The experiments are recorded
  in top-view by a steadily translating CMOS camera that is mounted
  onto a motorised translation stage. (b) The channel's
  cross-section. (c) The five-step procedure that is used to generate
  a centred bubble prior to an experiment.}
\label{fig:setup}
\end{figure}

The experiments were performed inside a rectangular Hele--Shaw channel
[Fig.~\ref{fig:setup}\textcolor{blue}{(a)}], which has previously been described comprehensively by \citet{LifeAndFate}. The channel consisted of two float glass plates that were separated by two parallel sheets of steel
shim [Fig.~\ref{fig:setup}\textcolor{blue}{(b)}]. The width of the
channel was $W^* =$ \SI{40.0 (1)}{\milli\metre} and its depth was $H^*
=$ \SI{1.00 (1)}{\milli\metre}. The channel was approximately 2 metres
long. The channel was filled with silicone oil (Basildon Chemicals
Ltd) of dynamic viscosity $\mu =$ \SI{0.019}{\pascal \second}, density
$\rho =$ \SI{951}{\kg \per \cubic \metre} and surface tension $\sigma
=$ \SI{21}{\milli \newton \per \metre} at the ambient laboratory
temperature of \SI{21(1)}{\celsius}. The flow of oil was controlled by
two syringe pumps (KD Scientific) that were connected together in
parallel with rigid plastic tubes. These withdrawal  syringe pumps were linked to the channel outlet and an external oil reservoir with a three-way solenoid valve; this allowed us to control flow into/out of the channel and refill the channel prior to experiments. The channel was connected to the atmosphere by a two-way air valve that was situated a short distance downstream of the channel inlet. The syringe pumps and solenoid valves were linked to TTL switches and controlled on a computer with custom-built LabVIEW programs.

We generated a bubble inside the channel by manually injecting air
through a small tube that was threaded through the channel inlet and
connected to a \SI{25}{\micro \litre} syringe; this method did not
allow us to precisely control the bubble's volume. However, the
bubble's size was measured precisely using an edge-detection algorithm
and repeat experiments were performed in order to span the required
range of bubble sizes. The bubble was propagated to a position
downstream of the air valve by withdrawing a small quantity of oil and
aligned approximately with the channel's centreline; this was achieved
non-invasively by tilting the channel about its streamwise axis and
allowing the bubble to free-rise under buoyancy
[Fig.~\ref{fig:setup}\textcolor{blue}{(c)}]. The channel was returned
to its horizontal position and the experiment was initiated by
withdrawing oil at a constant volumetric flow rate $Q^*$ whilst the
air valve was open to the atmosphere. The invading finger of air was
filmed in top-view by an overhead CMOS camera that was mounted onto a
motorised translation stage. The camera's observation window was $1760
\cross 330$ pixels and, depending on the value of $Q^*$, the camera's
frame rate was varied between $12$ and $60$ frames per second. The finger's propagation speed following aggregation with the bubble is not known a priori and, so, the camera's translation speed was manually adjusted until it moved at approximately the same speed as the finger's tip in order to keep it in the camera's field of view throughout the experiments. The air-oil interface was visualised by illuminating the channel from below with a custom-built LED light box; this led to the interface appearing as a darkened contour when viewed from above. However, this technique leads to small (quantifiable) uncertainties in our measurements because the interface appears to have a finite thickness of approximately 2 pixels. The contours were detected with a Canny edge-detection algorithm, implemented using a custom-built Python script, which allowed for measurements of the bubble's size, the finger's width and its propagation speed.

We will use the channel's half-width $W^*/2$ and the average speed of
oil $U_0^* = Q^* / (W^* H^*)$, respectively, as our characteristic
length and velocity scales. The two-dimensional coordinate system ($x$, $y$) is used to describe the finger, where $x = 2 x^* / W^*$ and $y = 2 y^* /
W^*$ are the dimensionless coordinates spanning the length and width
of the channel, respectively. The line $y=0$ corresponds to the axial
centreline of the channel, whilst the channel's side-walls are given by the lines $y = \pm 1$. The dimensionless flow rate is given by $Q
= \mu U_0^* / \sigma$, which can also be interpreted as a capillary
number based on the average speed of the oil, and takes values between $0 \leq Q \leq 0.12$ in our experiments. The finger's speed $U^*$ was determined by calculating the streamwise displacement of its tip across a series of consecutive frames, and its dimensionless speed relative to the mean speed of the surrounding oil is given by $U = U^* / U_0^*$. The bubble's size is parametrised in terms of a dimensionless radius $r = 2\sqrt{A^* / \pi} / W^*$, where $A^*$ is the projected area measured in-flow through image analysis, and we note that this differs from the stationary projected area due to three-dimensional (thin-film) and compressional effects \citep{LifeAndFate}.

\begin{figure}
\includegraphics[width=\textwidth, clip]{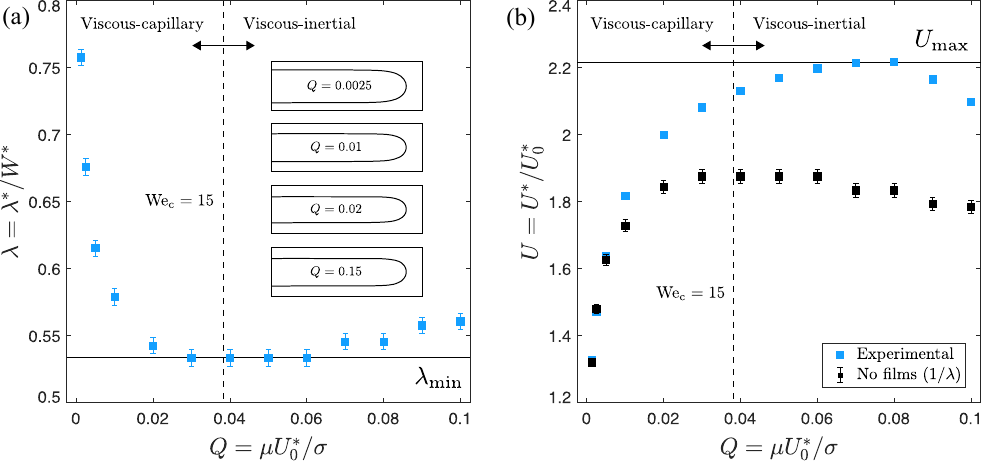}
\caption{(a) The fraction of the channel's width $\lambda = \lambda^* / W^*$ occupied by the Saffman--Taylor finger and (b) the dimensionless speed $U = U^* / U_0^*$ as the dimensionless flow rate $Q = \mu U_0^* / \sigma$ varies. The critical value of the modified Weber number $\mathrm{We}_c = \rho U_c^{*2} W^* / \sigma = 15$ that was identified by \citet{chevalier} is indicated by the dashed vertical line. The channel's width is $W^* =$ \SI{40.0 (1)}{\milli\metre} and its depth is $H^* =$ \SI{1.00 (1)}{\milli\metre}.}
\label{fig:ST_finger}
\end{figure}

\subsection{The Saffman--Taylor finger}
\label{sec:ST_finger}

\begin{figure}
\includegraphics[width=\textwidth, clip]{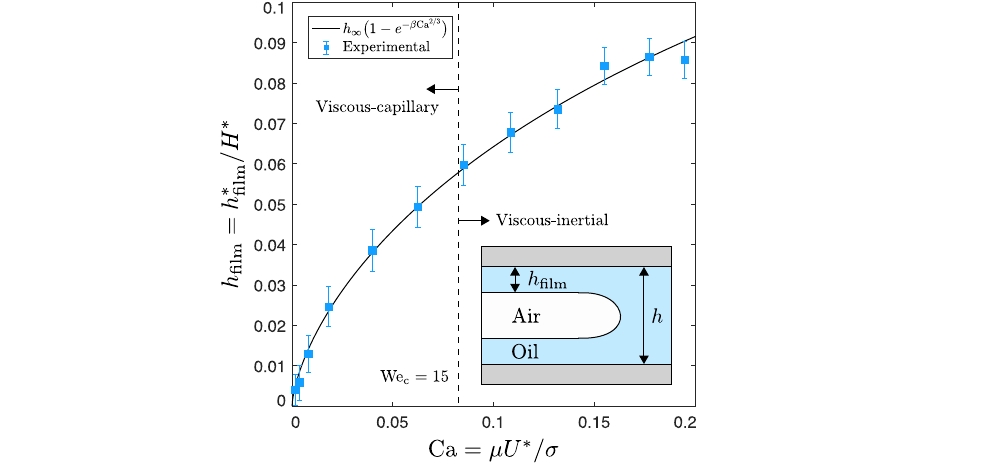}
\caption{Variation of the dimensionless mean liquid film thickness $h_{\textnormal{film}} = h_{\textnormal{film}}^* / H^*$ as a function of the capillary number $\mathrm{Ca} = \mu U^* / \sigma$. The experimental data points are well-described by a $\mathrm{Ca}^{2/3}$ power law relationship; see Eq.~(\ref{eqn:films2}). Inset: simplified side-view schematic diagram of the liquid films that are deposited above and below the finger as it propagates.}
\label{fig:film_thickness}
\end{figure}

The two plots in
\cref{fig:ST_finger} show (a) the fraction of the channel's width $\lambda = \lambda^* / W^*$ occupied by the Saffman--Taylor finger, and (b) the dimensionless speed $U = U^* / U_0^*$ as the dimensionless flow rate $Q = \mu U_0^* / \sigma$ varies. As the flow rate increases, the finger width initially decreases because the corresponding increase in viscous pressure drop leads to increased curvature of the finger tip. The half-width finger is not observed in our experiments and, instead, the finger's width plateaus at a minimum value $\lambda = \lambda_{\textnormal{min}}
\equiv 0.53$ before increasing slightly.
This departure from the depth-averaged theory is due to the influence of inertial
forces at higher propagation speeds, which tend to broaden the finger \citep{chevalier}. \citet{chevalier} found that the onset of ``inertial-broadening'' occurred universally at the critical Weber number $\mathrm{We} = \rho U^{*2} W^* / \sigma = 15$, which coincides with the observed plateau. The critical Weber number delineates the transition between the ``viscous-capillary'' regime, where the finger's width decreases monotonically with increasing flow rate, and the ``viscous-inertial'' regime, where the finger's width increases monotonically with increasing flow rate. However, although inertial effects affect the finger width at higher flow rate, we will present evidence to show that they do not play an important role in the observed oscillatory dynamics.

The finger's dimensionless speed is related to its relative width by mass conservation, leading to a simple relation $U = 1 / \lambda$ in the two-dimensional ($h \rightarrow 0$) limit. For finite $h$, the relationship between the finger's width and speed, see \cref{fig:ST_finger}(b), is modified by the presence of liquid films above and below the finger, arising from the wettability of the liquid with the channel's boundaries. At low $Q$, the films are thin and $U$ is approximately $1/\lambda$. At higher $Q$, the films increase in thickness and the finger effectively propagates into a channel of reduced height, which increases its propagation speed for fixed $\lambda$. 
The mean film thickness is plotted in Fig.~\ref{fig:film_thickness} as a function of the capillary number, and we find that the experimental measurements are well-described by the power law relationship
\begin{equation}
    h_{\textnormal{film}} = 0.19 \bigg{[}1-\exp (- 1.88 \: \mathrm{Ca}^{2/3})\bigg{]}
    \label{eqn:films2}
\end{equation}, which is consistent with the empirical result that was derived by \citet{tabeling}. Following the onset of inertial broadening, the finger's dimensionless speed continues to increase with increasing flow rate because the increasing finger width, which would cause a reduction in speed, is outweighed by the increasing film thickness. However, the converse is true for
higher flow rates, as $U$ decreases with
increasing $Q$ for $Q \geq 0.08$, see Fig.~\ref{fig:ST_finger}.


\begin{figure}
\includegraphics[width=\textwidth, clip]{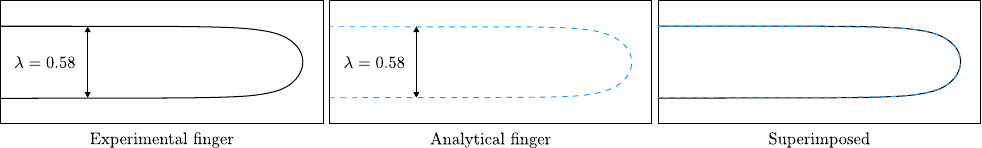}
\caption{The agreement between the experimental finger shape and analytical finger shape given by Eq.~(\ref{Pitts_profile_eqn}) for $\lambda = \lambda^* / W^* = 0.58$. The dimensionless flow rate is $Q=\mu U_0^* / \sigma = 0.01$.}
\label{fig:analytical}
\end{figure}

The finger shapes observed in experiments for $\lambda \leq 0.50$ are well-described by the analytical family of finger shapes found by \citet{Saffman1958}. For $\lambda > 0.50$, \citet{Pitts} used an empirical relationship for the local curvature at a point on the finger's interface to derive a modified family of finger shapes. A closed-form expression for the in-plane shape of a finger of relative width $\lambda = \lambda^* / W^*$ that combines these two results is given by:
\begin{equation}
    x(y; \lambda) = \frac{F(\lambda)}{\pi} \ln{} \bigg{[} \frac{1 + \cos(\pi y / \lambda)}{2} \bigg{]},
\label{Pitts_profile_eqn}
\end{equation}
where ($x = 2x^* / W^*, y = 2 y^* / W^*$) are the dimensionless
coordinates of the interface and the origin $(x, y) = (0, 0)$ is
situated at the finger's tip. The function $F(\lambda)$ is given by
$F(\lambda) = (1 - \lambda)$ for $\lambda \leq 0.50$ and $F(\lambda) =
\lambda$ for $\lambda > 0.50$. The agreement between experiments and
theory is demonstrated by the representative comparison in
Fig.~\ref{fig:analytical}; here we have overlaid the experimental and
equivalent-width analytical fingers. 



\section{Results}

\subsection{Phase diagram of the finger's long-term behaviour}
\label{phase_diagram}

\begin{figure}
\includegraphics[width=\textwidth, clip]{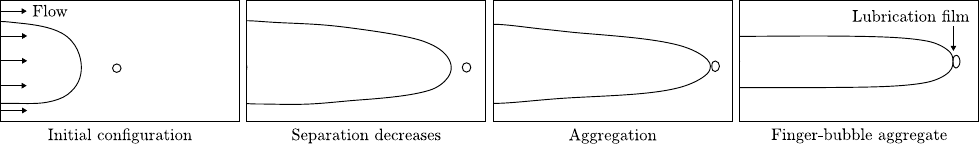}
\caption{The general behaviour of the system shortly after the finger is formed by withdrawing the oil at a constant flux from the opposing end of the channel whilst the air valve is open to the atmosphere. The finger catches up to the bubble and aggregates with it; immediate coalescence is prevented by a thin liquid film that separates the finger and the bubble. Post-aggregation, the finger-bubble aggregate either propagates steadily or unsteadily depending on the bubble's size and flow rate.}
\label{fig:general_behaviour}
\end{figure}

We will now investigate how the bubble influences the finger's
long-term mode of propagation. The time-sequence in
Fig.~\ref{fig:general_behaviour} shows the typical dynamics shortly
after the finger is formed. The finger initially catches up to the bubble because its
propagation speed is faster, and this leads to their aggregation. However, the coalescence between the bubble and finger is delayed because they are separated by a thin liquid film. For $Q < 0.02$, the liquid film ruptures shortly after aggregation, and this leads to coalescence; see movie 1 in the Supplemental Material. However, the liquid film persists throughout the entirety of the experiment for higher flow rates, irrespective of the bubble's size. The bubble, which deforms as the finger pushes up against its rear, replaces the finger's tip, and the system evolves until the finger-bubble aggregate assumes either a steady, periodic or continuously evolving mode of propagation.

\begin{figure}
\includegraphics[width=\textwidth, clip]{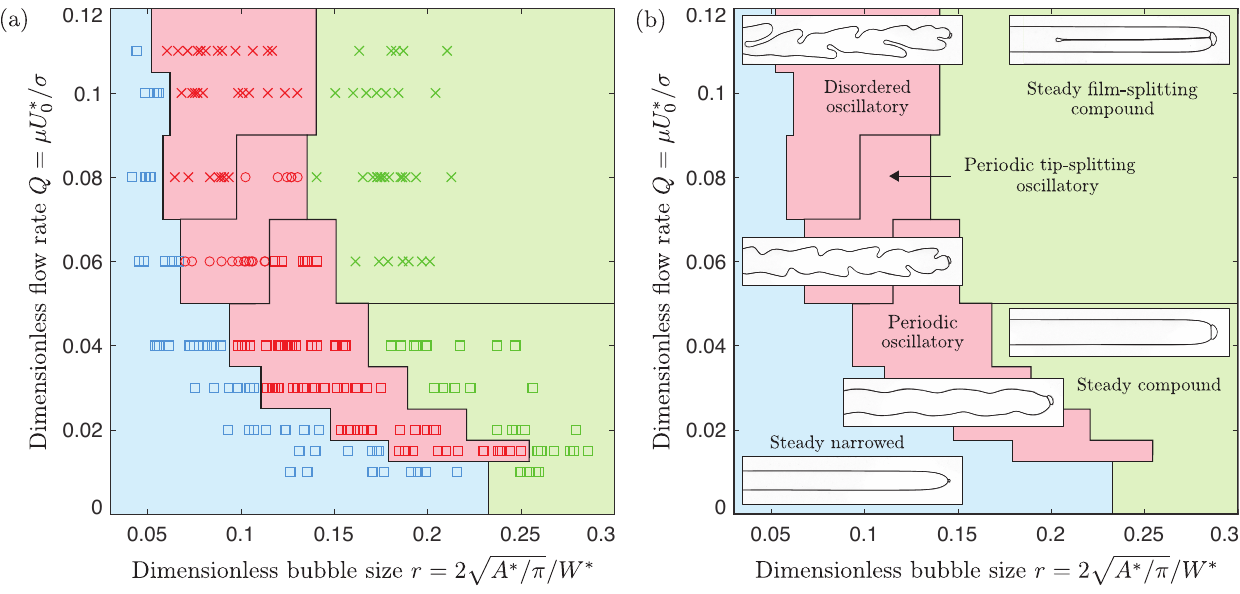}
\caption{(a) Two-dimensional phase diagram that characterises the influence of the bubble's dimensionless size $r = 2 \sqrt{A^* / \pi} / W^*$ (horizontal axis) and the dimensionless flow rate $Q = \mu U_0^* / \sigma$ (vertical axis) on the finger's long-term behaviour. The channel's width is $W^* =$ \SI{40.0 (1)}{\milli\metre} and its depth is $H^* =$ \SI{1.0 (1)}{\milli\metre}. We were unable to obtain measurements for $r < 0.05$ because the bubble was advected around the finger's tip and subsequently left behind instead of aggregating with it. This behaviour is attributed to the unavoidable lateral offset in the bubble's initial position that becomes important at small bubble sizes. The data markers are used to indicate different long-term fingering behaviours and they have been grouped more broadly into three coloured bands. The piecewise-linear solid lines are the approximate boundaries between regions of different long-term fingering behaviours. (b) Reproduction of the phase diagram with the data markers removed. The overlaid images are examples of the typical finger shapes. The movies 2-7 in the Supplemental Material are representative examples of these fingers.}
\label{fig:phase_diagram}
\end{figure} 

The two-dimensional phase diagram in Fig.~\ref{fig:phase_diagram}, which
contains the results of over two hundred experiments, characterises
the influence of the bubble's dimensionless size $r = 2\sqrt{A^* /
  \pi} / W^*$ (horizontal axis) and dimensionless flow rate $Q = \mu
U_0^* / \sigma$ (vertical axis) on the finger's long-term
behaviour. We identified six distinct long-term propagation modes over
the range of investigated parameters, occurring in distinct simply connected regions of the parameter space, and they have been
represented by different data markers; see movies 2-7 in the
Supplemental Material.
We have further grouped the long-term fingering behaviours that are qualitatively similar into three coloured regions corresponding to steadily propagating narrowed fingers, oscillatory fingers and steadily propagating compound fingers.

\subsubsection{Steadily propagating fingers}
\label{steadily}

In \cref{fig:phase_diagram} there are  two simply connected regions, which we have coloured in green and blue, of steadily propagating fingers. The fingers in these two regions both propagate symmetrically about the channel's centreline, but the fingers in the blue-coloured region, henceforth termed ``steadily propagating narrowed'', are characteristically narrower than the unperturbed Saffman--Taylor fingers; whereas, the fingers in the green-coloured region, henceforth termed ``steadily propagating compound''  are approximately the same width as the unperturbed Saffman--Taylor fingers. The two regions of steadily propagating fingers persist for all investigated flow rates.

\begin{figure}
\includegraphics[width=\textwidth, clip]{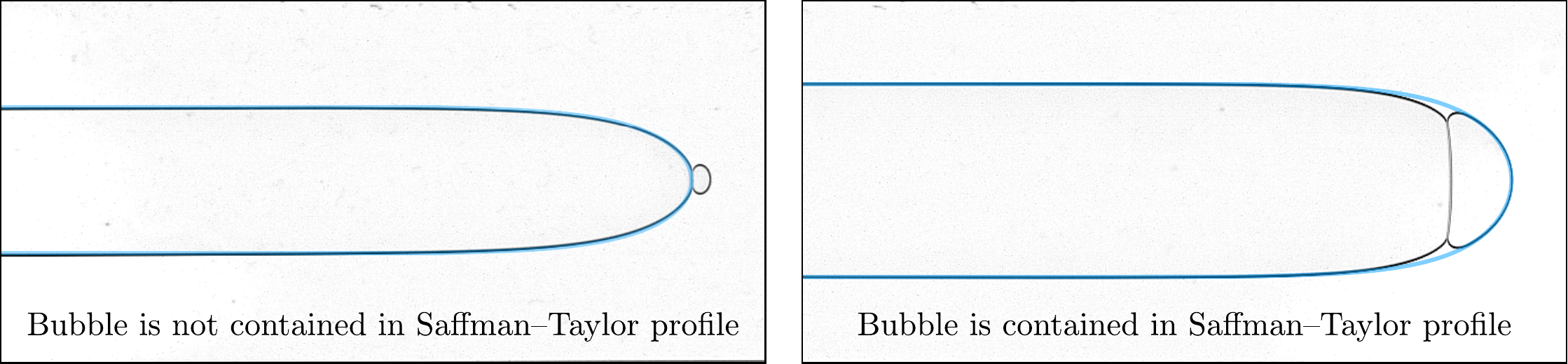}
\caption{The difference between the finger shapes in the two regions of steadily propagating fingers. The overlaid blue-coloured contours correspond to the analytical profiles of the equivalent-width Saffman--Taylor fingers. For steadily propagating narrowed fingers (the blue-coloured region in Fig.~\ref{fig:phase_diagram}), the finger's profile noninclusive of the bubble is well-described by the analytical family of Saffman--Taylor fingers. However, for steadily propagating compound fingers (the green-coloured region in Fig.~\ref{fig:phase_diagram}), the finger's profile inclusive of the bubble is well-described by the analytical family of Saffman--Taylor fingers.}
\label{fig:steady_fingers}
\end{figure} 

 Fig.~\ref{fig:steady_fingers} shows the equivalent-width Saffman--Taylor fingers (blue-coloured contours) from equation (\ref{Pitts_profile_eqn}), for two representative finger profiles. For steadily propagating narrowed fingers, the finger's profile noninclusive of the bubble is well-described by the analytical family of Saffman--Taylor fingers, which indicates that the bubbles are acting as localised flow field perturbations \citep{couder}. We suspect that the physical mechanism for the narrowing in this case is similar to how a rigid disk placed in front of the finger's tip can cause it to narrow \citep{disk} by locally reducing the fluid pressure, which increases the curvature of the finger tip. For steadily propagating compound fingers, the entirety of the finger's profile inclusive of the bubble is well-described by the analytical family of Saffman--Taylor fingers. The bubbles in this regime are large enough to effectively replace the finger's tip and become part of the overall finger shapes. Thus, despite the presence of the bubble at its tip, a steadily propagating finger adopts one of the the analytical family of Saffman--Taylor finger shapes.

\begin{figure}
\includegraphics[width=\textwidth, clip]{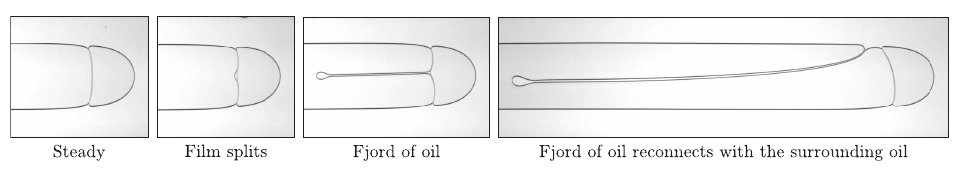}
\caption{The splitting of the liquid film between the bubble and finger that occurs for compound fingers at high flow rates. The splitting of the liquid film leads to the development of a narrow fjord of liquid inside the finger that progressively bends towards one of its sides.}
\label{fig:compound_split}
\end{figure}

The time-sequence in Fig.~\ref{fig:compound_split} shows an effect that occurs for steadily propagating compound fingers for $Q \geq 0.05 \pm 0.01$. Here, the liquid film that separates the bubble and finger splits, leading to the development of a narrow fjord of liquid inside the finger. The liquid fjord progressively bends towards one side of the finger, before connecting with the surrounding liquid as the finger advances. The connecting of the fjord with the surrounding
liquid leads to a slight ``wobbling'' of the finger tip as it propagates. The splitting of the liquid film occurs irregularly, and its initial location varies, indicating that this effect may be driven by perturbations. The fjord length depends on how symmetrically the liquid film splits, as longer fjords develop if the splitting begins close to the centre of the film, whilst shorter fjords develop if the splitting occurs in off-centred positions \citep{Kopf_Sill,lajeunesse}. We have not yet found an explanation for why this splitting occurs, but it appears as though the liquid inside the fjord originates from the liquid films that are left behind by the finger on the upper and lower channel boundaries. The interface at the finger tip is approximately flat, so the fjords may arise from a modification of the Saffman--Taylor instability, but the observed wavelength is shorter than the theoretical value of $\pi Ca^{-1/2}H^{*}$ \cite{Saffman1958}. 

\subsubsection{Oscillating fingers}

\begin{figure}
\includegraphics[scale=0.6]{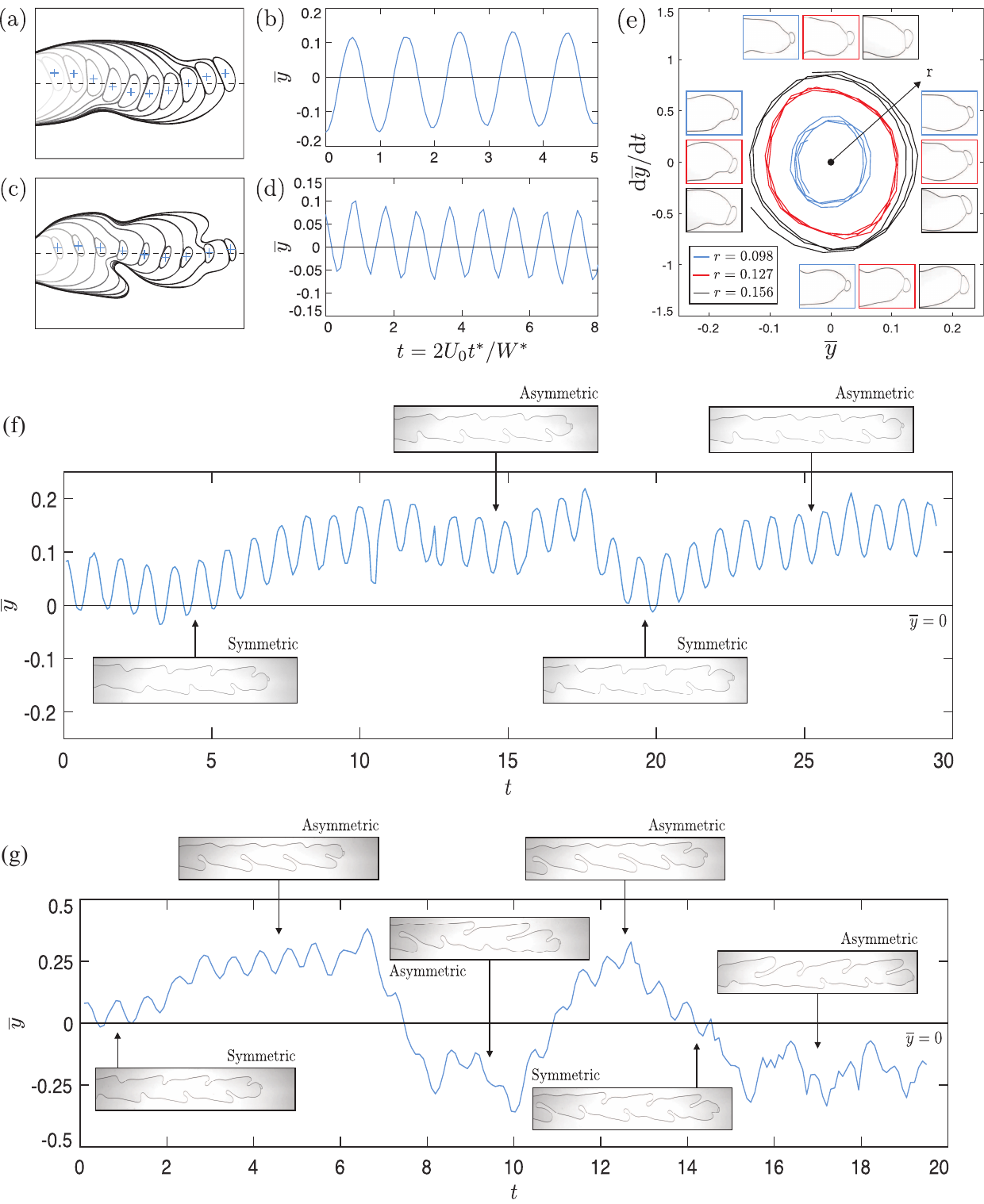}
\caption{(a) Composite image of a periodic oscillatory finger's front over a single cycle of oscillation. The blue-coloured markers indicate the position of the bubble's centroid. The bubble's size is $r = 0.156$ and the dimensionless flow rate is $Q = \mu U_0^* / \sigma = 0.04$. (b) Time-evolution of the $y$-coordinate of the bubble's centroid shown in (a). (c) Composite image of a periodic tip-splitting oscillatory finger's front over a single cycle of oscillation. The bubble's size is $r=0.102$ and the dimensionless flow rate is $Q = \mu U_0^* / \sigma = 0.08$. (d) Time-evolution of the $y$-coordinate of the bubble's centroid shown in (c). (e) Two-dimensional phase plane portrait of the finger's trajectories in the $(\overline{y}, \: \mathrm{d}\overline{y} / \mathrm{d} t)$ projection as the bubble's size increases. The steadily propagating narrowed fingers are stable fixed points situated at the origin $(0, \: 0)$ in this particular projection. Following the onset of periodic oscillations, the trajectories correspond to limit cycles centred about the origin. The overlaid images correspond to the finger's shape at the two minima and maxima of $\overline{y}$ and $\mathrm{d}\overline{y} / \mathrm{d} t$. The flow rate is $Q = \mu U_0^* / \sigma = 0.04$. (f) Time-evolution of the $y$-coordinate of the bubble's centroid. The insets show the finger's shape at various points in time. The bubble's size is $r=0.065$ and the dimensionless flow rate is $Q = \mu U_0^* / \sigma = 0.08$. (g) Time-evolution of the $y$-coordinate of the bubble's centroid. Error bars have not been included for visual clarity. The insets show the finger's shape at various points in time. The bubble's size is $r=0.067$ and the dimensionless flow rate is $Q = \mu U_0^* / \sigma = 0.10$.}
\label{fig:oscillations}
\end{figure} 

For $Q \leq 0.0125 \pm 0.0005$, the system transitions directly between steadily propagating narrowed fingers and steadily propagating compound fingers. However, for higher flow rates, the two regions of steadily propagating fingers are separated by the region of intermediate sized bubbles, coloured red in \cref{fig:phase_diagram}, in which the finger's tip oscillates laterally as it propagates. The region of oscillations is approximately constant in width, but shifts to smaller bubble sizes as the flow rate increases, suggesting a connection between bubble size and the unperturbed finger width which initially decreases with increasing flow rate, see Fig.~\ref{fig:ST_finger}. We identify three different types of oscillations, which generally increase in complexity as the flow rate increases: 

(1) For $Q \leq 0.05 \pm 0.01$, the finger's tip periodically meanders about the channel's centreline $y=0$ for all bubbles in the red-coloured region, leading to the development of regular ``wave-like'' fingering patterns [Fig.~\ref{fig:oscillations}\textcolor{blue}{(a)}]. The periodic oscillations of the finger's tip are reflected in the lateral displacement of the bubble's centroid from the channel's centreline $\overline{y} = 2 y^* / W^*$ in Fig.~\ref{fig:oscillations}\textcolor{blue}{(b)}. 

(2) For $0.05 \leq Q \leq 0.09$, the periodic oscillations are accompanied by secondary tip-splitting for a range of bubble sizes, leading to the development of increasingly complex patterns [Fig.~\ref{fig:oscillations}\textcolor{blue}{(c)}]. The growth of side-tips does not influence the motion of the finger's tip, as evidenced by the time-evolution of $\overline{y}$ in Fig.~\ref{fig:oscillations}\textcolor{blue}{(d)}. 

(3) For $Q > 0.09 \pm 0.01$, periodic oscillations are replaced by disordered dynamics, characterised by a seemingly random meandering of the finger's tip as it oscillates, see Fig.~\ref{fig:oscillations}\textcolor{blue}{(f,g)}. In fact, the disordered dynamics first appear for smaller bubbles in the red-coloured region at $Q = 0.07 \pm 0.01$, but they occur for all bubbles in the red-coloured region at $Q \geq 0.09 \pm 0.01$.

We have projected the bubble's trajectories at $Q = \mu U_0^* / \sigma = 0.04$ onto the two-dimensional $(\overline{y}, \mathrm{d}\overline{y} / \mathrm{d}t)$ plane in \cref{fig:oscillations}\textcolor{blue}{(e)}. The phase portrait is qualitatively representative of all flow rates in which the steadily propagating finger destabilises into stable periodic states. The steadily propagating fingers are stable fixed points lying at the origin $(0, \: 0)$ in this particular projection. Following the onset of oscillations, the phase-plane trajectories form closed loops (i.e. periodic orbits) centred about the origin $(0, \: 0)$. The amplitude of the periodic orbits increases as the bubble grows in size following the onset of oscillations, which is dynamically consistent with the existence of a supercritical Hopf bifurcation, in which the steadily propagating state destabilises at critical combinations of the control parameters. 

 For the higher flow rates, the irregular patterns left behind by the finger are qualitatively reminiscent of those observed following the destabilisation of the Saffman--Taylor finger at large values of $1/B$ \citep{tabeling}. The continuously evolving dynamics in this region are accompanied by high sensitivity, because repeat experiments with similar bubble sizes lead to strikingly different evolutions.
 
 The time-evolution of $\overline{y}$ shown in Fig.~\ref{fig:oscillations}\textcolor{blue}{(f)} is a representative example of the disordered dynamics at the lowest flow rate for which they occur $Q = \mu U_0^* / \sigma = 0.08$. The finger's tip oscillates with a well-defined period, but, unlike oscillations at lower flow rates, the tip no longer oscillates about the channel's centreline. The complex finger-tip oscillations do have an underlying structure, however, and appear to drift regularly between a symmetric (centred) mode and an asymmetric mode. 

The finger's dynamics, and consequently the generated patterns, become increasingly complex at higher flow rates. The time-evolution of $\overline{y}$ in Fig.~\ref{fig:oscillations}\textcolor{blue}{(g)} is a representative example of the disordered dynamics at the highest flow rate investigated $Q = \mu U_0^* / \sigma = 0.10$. The time-evolution now features abrupt switches, as well as slower drifts reminiscent of those described previously, between different types of oscillations. 

The two time-evolutions shown in Fig.~\ref{fig:oscillations}\textcolor{blue}{(f,g)}, although different in their complexity, are both dynamically consistent with the transient exploration of multiple, unstable periodic states. The symmetric oscillations are similar to the original periodic states, but they are no longer stable; and the asymmetric oscillations appear to be new periodic states that were not observed at lower flow rates.
In Fig.~\ref{fig:oscillations}\textcolor{blue}{(f)}, the behaviour is consistent with the interaction of two periodic states with different frequencies; at higher flow rates, Fig.~\ref{fig:oscillations}\textcolor{blue}{(g)}, the behaviour is consistent with interaction of more than two periodic states with different frequencies.
In the latter case, the development of large side-tips on the side of the finger furthest from the neighbouring side-wall in the asymmetric oscillations provides a physical mechanism to keep the finger off-centre, because they will tend to impede the finger's movement towards the channel's centreline. The asymmetric oscillations on the bottom side of the channel are irregular, unlike those on the upper side of the channel which are nearly perfectly periodic, and we speculate that this behaviour is driven by asymmetries (e.g. the unavoidable bias in the levelling of the channel) in experiments.


\subsubsection{Relative finger width and dimensionless speed}
\label{sec:finger_width}

\begin{figure}
\includegraphics[width=\textwidth, clip]{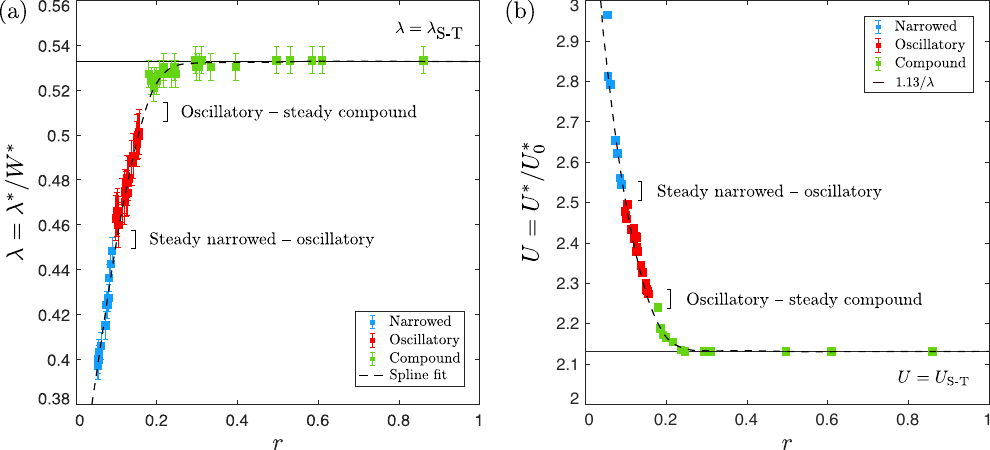}
\caption{Variation of the finger's (a) relative width $\lambda =
  \lambda^* / W^*$ and (b) dimensionless speed $U = U^* / U_0^*$ as
  functions of the bubble's size $r$ at $Q = \mu U_0^* / \sigma =
  0.04$. The blue data markers correspond to steadily propagating
  narrowed fingers, red data markers correspond to oscillatory fingers,
  and green markers correspond to steadily propagating compound
  fingers. The width of steady fingers was determined from direct
  measurement, whilst the mean width of oscillatory fingers was
  determined indirectly from mass conservation by measuring its speed
  and interpolating the thickness of the deposited oil films. The
  solid curve in (b) shows that the finger's speed is inversely
  proportional to its width, which indicates negligible
  three-dimensional effects. The unperturbed finger's relative width
  and dimensionless speed are $\lambda_{ST} = 0.53$ and $U_{ST} =
  2.13$. The error bars in (a) arise from the apparent thickness of
  the air-oil interface, whilst error bars have been omitted in (b)
  because they are smaller than the data markers. The black dashed line indicates the curve $1.13/\lambda$ based on a spline fit through the experimental data points in (a) for $\lambda$.}
\label{fig:width_and_speed}
\end{figure}

 The two plots in Fig.~\ref{fig:width_and_speed} show the finger's (a) relative width $\lambda = \lambda^* / W^*$ and (b) dimensionless speed $U = U^* / U_0^*$ as functions of the bubble's size at $Q = \mu U_0^* / \sigma = 0.04$, and they are qualitatively representative of all investigated flow rates. 
 
  As the bubble width increases, the finger width increases and the finger speed decreases; the finger speed is approximately equal to $1.13/\lambda$. The increase in finger width occurs because the curvature imposed by the bubble at the finger tip decreases with increasing bubble size \citep{couder}: smaller curvature leads to a smaller pressure jump across the bubble tip, increasing the local fluid pressure and decreasing the overall fluid pressure gradient, which causes the system to select a slower, and therefore, broader finger. 
 
  For sufficiently large bubbles, the width and speed remain approximately constant with values corresponding to those of the single Saffman--Taylor finger at the imposed flow rate, indicating that once it is large enough the bubble acts as the tip of a steadily propagating compound finger.

The two plots in Fig.~\ref{fig:width_and_speed} also reveal characteristic features of the finger at the two transition points. The destabilisation of the steadily propagating narrowed finger, occurring at $r = 0.093 \pm 0.005$, characteristically occurs when the finger is significantly narrower than the unperturbed Saffman--Taylor finger. However, the onset of periodic oscillations does not influence the finger's propagation speed because we were unable to identify any unsteadiness (within experimental resolution) in time-evolutions of the finger's speed for oscillating fingers. The finger's mean width and speed continue to smoothly increase and decrease, respectively, as the bubble's size increases following the onset of oscillations. Based on these two observations, we infer that the finger's width is predominantly set by the unstable steadily propagating state from which the stable limit cycle emanates. The second transition occurs at $r = 0.167 \pm 0.009$, as the periodic oscillations cease and the finger returns back to steady propagation. Notably, this transition characteristically occurs when the finger's width is approximately equal to that of the unperturbed Saffman--Taylor finger. The relevance of this observation will later become apparent in Sec.~\ref{osc_to_compound} when describing the transition between oscillating and steadily propagating compound fingers.

The capillary number $\mathrm{Ca} = \mu U^* / \sigma \equiv QU$ determines the thickness of the liquid films above and below the air finger, and varies between $0.09 \leq \mathrm{Ca} \leq 0.12$ over the resulting range of finger speeds. The liquid films have an approximately constant thickness ($0.06 \leq h \leq 0.07$) in this range of capillary numbers [Fig.~\ref{fig:film_thickness}], which means that three-dimensional effects are not expected to play a significant role; a finding reinforced by the inversely proportional relationship between the finger's width and propagation speed, see Fig.~\ref{fig:width_and_speed}. It follows that the perturbation applied at the tip of the finger by the bubble is essentially a two-dimensional effect, which means that we can simplify our proceeding analysis by only considering the ``in-plane'' finger shapes.

\subsection{Transition to oscillations}
\label{oscillations}

\begin{figure}
\includegraphics[width=\textwidth, clip]{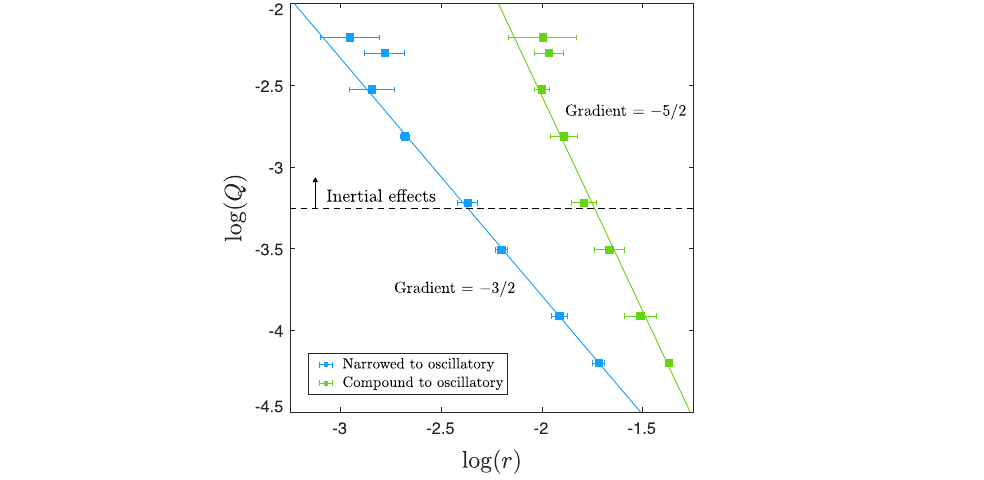}
\caption{Log-log plot of the two transition points in the ($r$, $Q$) plane. The experimental measurements are well-described by linear least-squares regression fits. The horizontal dashed line indicates where inertial effects cause the Saffman--Taylor finger to broaden as the flow rate increases.}
\label{fig:transitions}
\end{figure} 

The two transitions between steady and oscillatory behaviour shown in Fig.~\ref{fig:phase_diagram} are well-described by smooth $r^{\alpha}$ relationships over the investigated range of parameters, as indicated by the linear relationships between the values of $r$ and $Q$ on log-log axes in Fig.~\ref{fig:transitions}. These relationships indicate that the critical bubble radius depends on the geometry of the finger. In the absence of inertia, $\lambda \sim Ca^{-2/3}$ near the limiting finger width \citep{couder}. Hence, $Ca = UQ \sim \lambda^{-3/2}$, and if we assume that the finger width scales with bubble radius, $\lambda \sim r$, then $UQ \sim r^{-3/2}$. If $U$ remains approximately constant then the system is driven by the finger speed and we obtain the scaling $Q \sim r^{-3/2}$, which describes the transition from steady narrowed fingers to oscillations. Alternatively, if the system is driven by the bubble speed, then $U \sim r$ and we obtain the scaling $Q \sim r^{-5/2}$, which describes the transition from the compound finger to oscillations. The scaling $U\sim r$ can be found by balancing the viscous and capillary pressure drops over an isolated bubble.
We note that these scalings do not explain what sets the critical bubble radius for each transition, but they do indicate how such a critical radius varies with flow rate.

The observed deviation from these scalings at high values of $Q$ coincides with the decrease in finger propagation speed due to inertial effects, see  Sec.~\ref{sec:ST_finger}. Thus, although inertial effects cannot be neglected for sufficiently high flow rates, they do not appear to play a significant role in the transitions to oscillatory behaviour over the investigated range of parameters, indicating that the mechanism for oscillations is not inertially driven.

\begin{figure}
\includegraphics[width=\textwidth, clip]{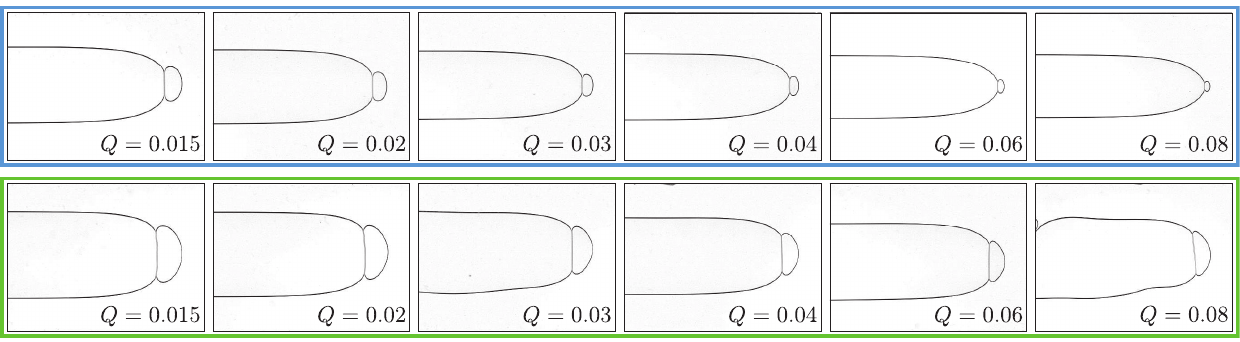}
\caption{The limiting steadily propagating finger shapes on both sides of the region of oscillations at different flow rates $Q = \mu U_0^* / \sigma$. The top row of images correspond to the largest bubbles that resulted in steadily propagating narrowed fingers whilst the bottom row of images correspond to the smallest bubbles that resulted in compound fingers. The compound fingers exhibit slight unsteadiness at high flow rates because the liquid film that separates the bubble and finger repeatedly splits.}
\label{fig:transition_shapes}
\end{figure}

The two sets of images in Fig.~{\ref{fig:transition_shapes}} show the limiting steadily propagating finger shapes bounding either side of the region of oscillations for different flow rates. We conjecture that the key to understanding the transition mechanisms lies in identifying common geometrical properties linking these two sets of steadily propagating finger shapes, and we will now examine how the steadily propagating finger and bubble shapes change leading up to the onset of oscillations.

\subsubsection{Narrowed to oscillatory transition}
\begin{figure}
\includegraphics[width=\textwidth, clip]{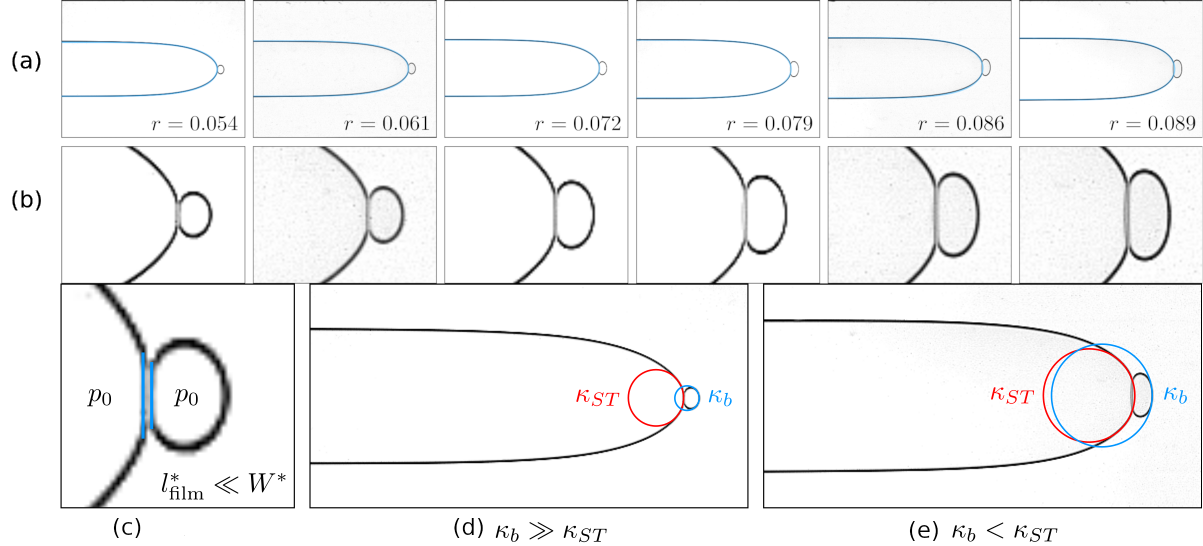}

\ \vspace*{-0.2cm}

\caption{The shapes of the steadily propagating narrowed fingers as the bubble's size increases towards the transition point at $Q = \mu U_0^* / \sigma = 0.04$. The top row of images (a) show the finger's overall shape whilst the middle row of images (b) are zoomed-in on the corresponding bubble's shape. The finger broadens marginally and the bubble's shape evolves from approximately circular to ``kidney bean''-like. The liquid film that separates the bubble and finger is straight within experimental resolution in all cases. The overlaid blue-coloured contours in the top row of images correspond to the analytical profiles of the equivalent-width Saffman--Taylor fingers. (c) The liquid film between the bubble and finger is straight and has negligible thickness. (d, e) The variation of the interface's curvature in the vicinity of the finger's tip. The regions of low curvature correspond to regions of high liquid pressure and vice-versa. The coloured circles indicate the osculating circles used to determine the bubble tip's curvature, $\kappa_b$. (d) Example in which the bubble has a higher tip-curvature $\kappa_b$ than the equivalent-width Saffman--Taylor finger's tip-curvature $\kappa_{ST}$. (e) Example in which the bubble has a lower tip-curvature $\kappa_b$ than the equivalent-width Saffman--Taylor finger's tip-curvature $\kappa_{ST}$.}
\label{fig:narrowed_transition_images}
\end{figure}

We start with the steadily propagating fingers resulting from the smallest bubbles and increase the bubble's size until the finger oscillates. The sequence of images in Fig.~\ref{fig:narrowed_transition_images} show the steadily propagating narrowed fingers and enlargements of the corresponding bubble shapes as the bubble's size increases at $Q = \mu U_0^* / \sigma = 0.04$. The overlaid blue-coloured contours are the analytical profiles of the equivalent-width Saffman--Taylor fingers and, as described in Sec.~\ref{phase_diagram}, they match the experimental finger profiles noninclusive of the bubble remarkably well in all cases. The finger's width is approximately constant ($0.40 \leq \lambda \leq 0.45$) over the investigated range of bubble sizes.

The shapes of the smallest bubbles are dominated by surface tension and are near circular. The shapes of the larger bubbles are influenced by viscous forces and are ``kidney bean-like'', notably different from the typical bubble shapes in Hele-Shaw cells, which tend to elongate in the direction of motion in order to reduce their resistance in viscous-dominated flow regimes; see \citet{monnet}. The observed lateral elongation is a consequence of the presence of the finger, which impedes the bubble's ability to elongate in the direction of motion.

\begin{figure}
\flushleft\hspace*{-0.5cm}\includegraphics[scale=0.8]{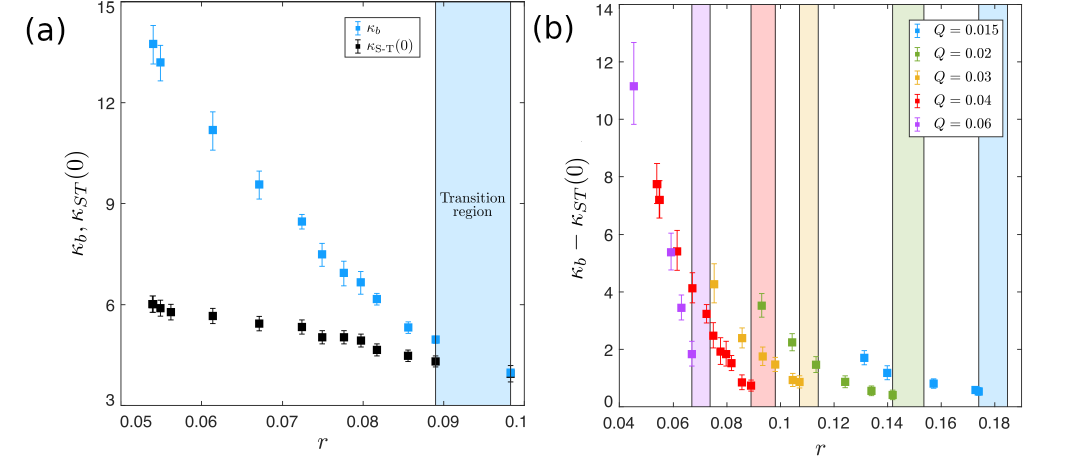}
\caption{(a) Variation of the bubble's tip-curvature and the equivalent-width Saffman--Taylor finger's tip-curvature as functions of the bubble's size $r$ at $Q = \mu U_0^* / \sigma = 0.04$. The bubble's tip-curvature was obtained by measuring the radius of its osculating circle. The equivalent-width Saffman--Taylor finger's tip-curvature was obtained by differentiating the analytical expression of its shape; see Eq.~(\ref{Pitts_profile_eqn}). The blue-coloured ``transition region'' represents the range of bubble sizes between the final steadily propagating and first oscillatory data points. The error bars arise from the apparent thickness of the air-liquid interface when analysing the recorded images.
(b) Variation of the difference between the bubble's tip-curvature $\kappa_b$ and the equivalent-width Saffman--Taylor finger's tip-curvature $\kappa_{\textnormal{S-T}}(0)$ as functions of the bubble's size $r$ at five different flow rates. The transition regions are indicated by the corresponding-coloured bands of bubble sizes. The error bars arise from the apparent thickness of the air-liquid interface in the captured frames.}
\label{fig:narrowed_transition}
\end{figure}

 The front of the bubble is approximately uniformly curved, which means that we can obtain the dimensionless curvature of its tip $\kappa_b = \kappa_b^* W^* / 2$ by measuring the radius $r_b = 2 r_b^*/W^*$ of its osculating circle, see Fig.~\ref{fig:narrowed_transition}(b,c). The curvature at the finger tip is found by using the analytic expression for the equivalent-width Saffman--Tayor finger, equation (\ref{Pitts_profile_eqn}), in the formula for curvature 
 $\kappa_{ST}(y;\lambda) = x'' / (1+x'^{2})^{3/2}$ evaluated at the tip:
\begin{equation}
    \kappa_{ST}(0; \lambda) = \left |{\frac{(\lambda - 1) \pi}{2 \lambda^2}} \right |.
\label{eqn:curvature}
\end{equation}

The two curvatures are plotted as functions of the bubble's size in Fig.~\ref{fig:narrowed_transition}(a) and the coloured ``transition region'' represents the interval of bubble sizes between the final steadily propagating and first oscillatory data points.  The curvature of the bubble's tip is initially higher than that of the finger's tip and both decrease as the bubble increases in size.
The difference between the two curvatures decreases sharply as the bubble's size increases, until they converge at the onset of oscillations.


When varying the flow rate, we consistently observed that the curvature of the bubble's tip converges towards that of the finger region as the bubble's size increases. The difference between the two curvatures is plotted in Fig.~\ref{fig:narrowed_transition}(b) as a function of the bubble's size for different flow rates, where the transition regions are represented by the corresponding-coloured bands. Hence, equality between the two curvatures is the limiting geometrical configuration for steadily propagating narrowed fingers.

As noted above, the liquid film thickness above and below the finger and bubble is approximately constant and so the interface curvature in the out-of-plane direction is also approximately constant. Hence, changes in fluid pressure can be inferred from the in-plane curvature of the interfaces. We note that the interfaces in the near-contact region between finger and bubble are approximately flat, see Fig.~\ref{fig:narrowed_transition_images}, meaning that the bubble pressure is approximately the same as the finger pressure.

For steadily propagating narrowed fingers, the maximum curvature of the interface and hence minimum fluid pressure is at the bubble's tip, which leads to a pressure gradient that propagates the finger forwards along the centreline of the channel. The enhanced pressure minimum in the vicinity of the finger's tip may be why \citet{couder} found that narrower fingers destabilise at higher values of $1/B$ compared to half-width fingers. 

Once the finger and bubble tip curvatures are the same, there is no axial pressure gradient across the bubble and it is susceptible to lateral perturbations. If displaced laterally both a sufficiently large isolated bubble and the finger will return to the channel's centreline, but they will do so on different timescales.  We conjecture that this difference in restoral timescales is what allows sustained oscillations to develop in the finger-bubble system. Fig.~\ref{fig:oscillations} shows that the laterally displaced bubble's shape adjusts to drive it back towards the channel centreline, but the bubble is then carried past the centreline by the restoring motion of the finger tip and the process begins again.

If the bubble is small then it does not deform; and without the presence of the finger it would not return to the channel's centreline after lateral displacement. The lack of an independent restoral mechanism for the bubble ensures the stability of the narrowed-finger state because the bubble passively follows the finger's restoral mechanism. This argument is supported by the scalings discussed in \S \ref{oscillations}, which reveal that the system is driven by the finger at the narrowed to oscillatory transition. 




\subsubsection{Oscillatory to compound transition}
\label{osc_to_compound}

\begin{figure}
\includegraphics[width=\textwidth, clip]{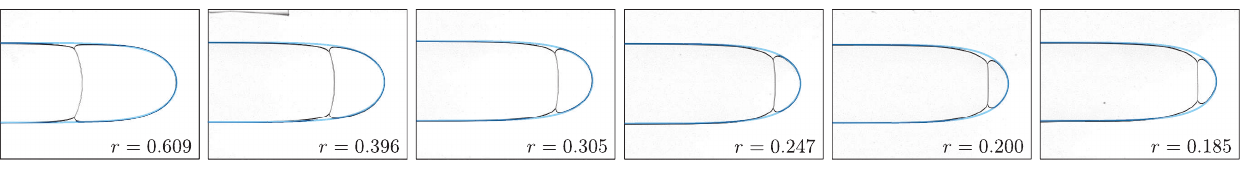}
\caption{The steadily propagating compound finger shapes as the bubble's size decreases towards the second transition point at $Q = \mu U_0^* / \sigma = 0.04$. The finger's width is approximately constant. The curvature of the liquid film that separates the bubble and finger decreases as the bubble's size decreases. The overlaid blue-coloured contours correspond to the analytical profiles of the equivalent-width Saffman--Taylor fingers.}
\label{fig:compound_shapes}
\end{figure}

We investigate the transition to oscillatory dynamics for larger bubbles by starting with the largest bubbles and progressively decreasing the bubble's size until the finger oscillates, see Fig.~\ref{fig:compound_shapes}. In the compound fingers, all bubbles adopt the shape of the tip of a single Saffman--Taylor finger. This means that the bubble's restoral timescale will be approximately the same as that of the finger and there is no longer a mechanism for sustained oscillation. Again, this argument is supported by the previously discussed scalings which indicate that the system is driven by the bubble at the compound to oscillatory transition.

For the largest bubbles, the viscous pressure drop associated with the equivalent single propagating finger takes place over the length of the bubble. In order for the finger to propagate there must be a pressure drop across the finger tip, which can only be induced by curvature of the finger tip. As the bubble size decreases, more of the viscous pressure drop takes place along the finger and the tip curvature decreases accordingly. The limiting configuration is one in which the finger-bubble interface region is approximately flat. Note that the bubble in the limiting configuration is not large enough to act as a Saffman--Taylor finger in isolation; it is only the additional deformation caused by interaction with the finger that enables the bubble to adopt the shape of the finger tip. Thus, the oscillations begin when the bubble is sufficiently small that it does not adopt the shape of the finger tip and its restoral timescale differs from that of the finger. Precisely what sets this threshold is not clear.



\begin{figure}
\includegraphics[width=\textwidth, clip]{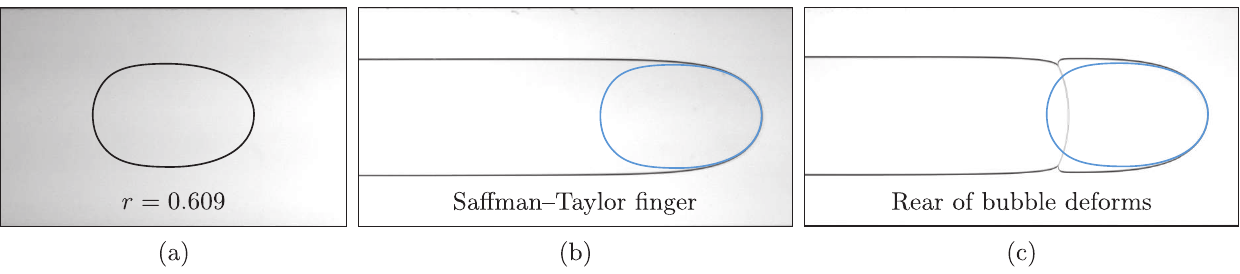}
\caption{(a) A steadily propagating bubble of size $r = 0.609$ at dimensionless flow rate $Q = \mu U_0^* / \sigma = 0.04$. (b) The bubble overlaid with the tip of the Saffman--Taylor finger. (c) The bubble overlaid with the steadily propagating compound finger, showing how the rear of the bubble deforms.}
\label{fig:big_compound}
\end{figure}

\section{Discussion}
\label{conclusion}

We have investigated experimentally the behaviour of Saffman--Taylor
fingers whose tips have been perturbed by bubbles. The experiments serve as an extension to those previously carried out by \citet{couder}, who
explored the anomalous selection of steadily propagating narrowed
fingers. The authors showed preliminary experiments of the anomalous
fingers destabilising into periodic states, yet provided no physical explanation for why this behaviour occurs. We are interested in
periodic fingering dynamics because we hypothesise that unstable periodic states are fundamental to the propensity of Saffman--Taylor fingers to exhibit disordered dynamics \citep{park_homsy}.

The perturbed fingers exhibit a multitude of different long-term behaviours over the investigated range of parameters. We constructed a phase diagram, which classifies
the regions of the system's parameter space in which the observed behaviours
occur, and found that the dynamics are well-organised. For fixed flow rates, there are two regions of steadily propagating fingers, which occur for the smallest and largest bubbles. The
smallest bubbles effectively act as flow field perturbations in the vicinity of the finger's tip, leading to the selection of narrower fingers. The largest bubbles, in contrast, effectively replace the finger's tip leading to compound fingers with the same widths as would be selected without the bubble. The observations
indicate that despite the presence of the bubble at its tip, the
finger always adopts a shape that belongs to the analytical family of Saffman--Taylor fingers, as indicated by \citet{couder}. The region of oscillations, which occurs for an interval of intermediately sized bubbles, is bounded by the two regions of steadily propagating fingers. Depending on the flow rate and bubble size, the oscillations may either be periodic or disordered. For low flow rates, the finger's tip exhibits periodic oscillations centred about the channel's centreline, whilst secondary tip-splitting events can accompany the oscillations at higher flow rates. The onset of periodic oscillations is dynamically consistent with a supercritical Hopf bifurcation, which occurs for critical combinations of the system's control parameters.

We established the underlying physical mechanism that destabilises the steadily propagating narrowed finger by examining the curvature of the interface in the vicinity of the finger's tip as the bubble's size varies. We determined empirically
that the steadily propagating narrowed finger destabilises when the curvature of the bubble's tip becomes smaller than that of the finger tip situated behind it, with this criterion occurring robustly for all investigated flow rates. When the curvatures are equal there is no axial pressure drop over the bubble and it is susceptible to lateral perturbations. The sideways movement of the finger couples with the deformability of the bubble, which provides a restoring mechanism to return the bubble back to the centreline of the channel, but the finger and bubble restoral mechanisms occur on different timescales leading to the oscillations. The finger returns to steady propagation once the bubble has become sufficiently large that it effectively replaces the finger's tip and becomes part of the overall finger shapes.

We have also shown preliminary evidence of the finger exhibiting disordered dynamics at the highest investigated flow rates. The finger's irregular motion in this regime appears to be orchestrated by unstable periodic states, as evidenced by short-lived periodic oscillations. We observed centred oscillations, which are identical to the stable periodic oscillations described at lower flow rates; and off-centred oscillations, in which the mean position of the finger's tip is biased towards one of the channel's side-walls. The disordered dynamics that we have observed are an interesting fluid mechanical phenomenon in their own regard and, importantly, are consistent with our proposed hypothesis that periodic states underpin the complex dynamics. However, because the periodic states are presumably unstable, this interpretation is purely speculative in the absence of a mathematical model and numerical simulations.

\begin{figure}
\begin{center}
\includegraphics[width=0.5\textwidth, clip]{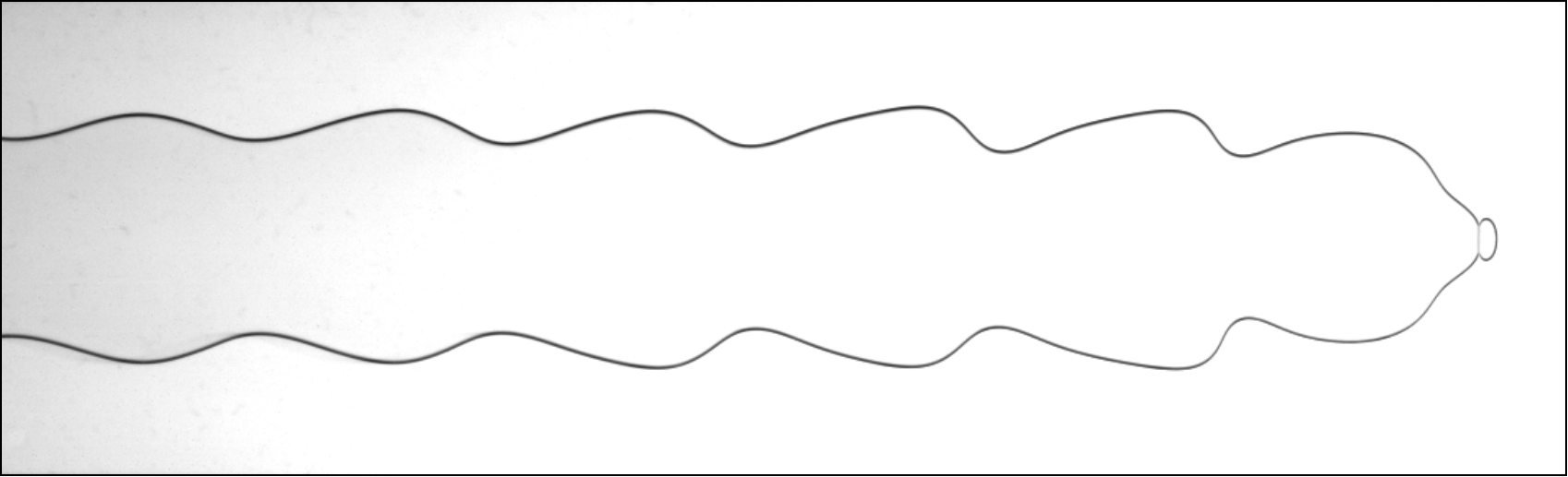}
\end{center}
\caption{The periodically pulsating mode of propagation that was observed in one of our experiments. The bubble's size is $r=0.054$ and the dimensionless flow rate is $Q = \mu U_0^* / \sigma = 0.10$.}
\label{fig:pulsating}
\end{figure}

The original experiments by \citet{couder} included preliminary observations of a different type of periodic oscillation, in which the bubble's shape ``pulsates'' symmetrically as the finger propagates. We only observed this mode of oscillation once in over two hundred experiments, occurring for a bubble that was slightly larger than the smallest bubble that we were able to generate in our experiments [Fig.~\ref{fig:pulsating}]. The fact that this state did not appear regularly raises the possibility that the pulsating oscillations arise from dynamically disconnected periodic states, which can be initiated by sufficient finite-amplitude perturbations. However, further experiments are required in order to clarify this.

The phenomena that we have observed further highlight the inherent complexity contained in this simple system, whilst the propensity of this system to exhibit periodic dynamics renders it a suitable candidate for studying the transition to disordered dynamics. Future modelling attempts may be aided by the observation that the liquid films play a negligible role in the finger's oscillations, because this suggests that two-dimensional models may be sufficient to capture the observed behaviours. However, the question still remains as to whether similar mechanisms are responsible for a single Saffman--Taylor finger's post-transition dynamics in the absence of a leading bubble.

\bibliographystyle{jfm}
\bibliography{bib.bib}

\end{document}